
\def\unredoffs{} 

\newbox\leftpage \newdimen\fullhsize \newdimen\hstitle \newdimen\hsbody
\tolerance=1000\hfuzz=2pt
\catcode`\@=11 
%
\magnification=1200\unredoffs\baselineskip=16pt plus 2pt minus 1pt
\hsbody=\hsize \hstitle=\hsize 
%
%
%
\newcount\yearltd\yearltd=\year\advance\yearltd by -1900

%
%

\def\draftmode{\message{ DRAFTMODE }\def\draftdate{{\rm preliminary draft:
\number\month/\number\day/\number\yearltd\ \ \hourmin}}%
\headline={\hfil\draftdate}\writelabels\baselineskip=20pt plus 2pt minus 2pt
 {\count255=\time\divide\count255 by 60 \xdef\hourmin{\number\count255}
  \multiply\count255 by-60\advance\count255 by\time
  \xdef\hourmin{\hourmin:\ifnum\count255<10 0\fi\the\count255}}}
\def\nolabels{\def\wrlabeL##1{}\def\eqlabeL##1{}\def\reflabeL##1{}}
\def\writelabels{\def\wrlabeL##1{\leavevmode\vadjust{\rlap{\smash%
{\line{{\escapechar=` \hfill\rlap{\sevenrm\hskip.03in\string##1}}}}}}}%
\def\eqlabeL##1{{\escapechar-1\rlap{\sevenrm\hskip.05in\string##1}}}%
\def\reflabeL##1{\noexpand\llap{\noexpand\sevenrm\string\string\string##1}}}
\nolabels
%
\global\newcount\secno \global\secno=0
\global\newcount\meqno \global\meqno=1
\def\newsec#1{\global\advance\secno by1\message{(\the\secno. #1)}
\global\subsecno=0\eqnres@t\noindent{\bf\the\secno. #1}
\writetoca{{\secsym} {#1}}\par\nobreak\medskip\nobreak}
\def\eqnres@t{\xdef\secsym{\the\secno.}\global\meqno=1\bigbreak\bigskip}
\def\sequentialequations{\def\eqnres@t{\bigbreak}}\xdef\secsym{}
\global\newcount\subsecno \global\subsecno=0
\def\subsec#1{\global\advance\subsecno by1\message{(\secsym\the\subsecno. #1)}
\ifnum\lastpenalty>9000\else\bigbreak\fi
\noindent{\it\secsym\the\subsecno. #1}\writetoca{\string\quad
{\secsym\the\subsecno.} {#1}}\par\nobreak\medskip\nobreak}
\def\appendix#1#2{\global\meqno=1\global\subsecno=0\xdef\secsym{\hbox{#1.}}
\bigbreak\bigskip\noindent{\bf Appendix #1. #2}\message{(#1. #2)}
\writetoca{Appendix {#1.} {#2}}\par\nobreak\medskip\nobreak}
%
%
\def\eqnn#1{\xdef #1{(\secsym\the\meqno)}\writedef{#1\leftbracket#1}%
\global\advance\meqno by1\wrlabeL#1}
\def\eqna#1{\xdef #1##1{\hbox{$(\secsym\the\meqno##1)$}}
\writedef{#1\numbersign1\leftbracket#1{\numbersign1}}%
\global\advance\meqno by1\wrlabeL{#1$\{\}$}}
\def\eqn#1#2{\xdef #1{(\secsym\the\meqno)}\writedef{#1\leftbracket#1}%
\global\advance\meqno by1$$#2\eqno#1\eqlabeL#1$$}
%
\newskip\footskip\footskip14pt plus 1pt minus 1pt 
\def\footnotefont{\ninepoint}\def\f@t#1{\footnotefont #1\@foot}
\def\f@@t{\baselineskip\footskip\bgroup\footnotefont\aftergroup\@foot\let\next}
\setbox\strutbox=\hbox{\vrule height9.5pt depth4.5pt width0pt}
\global\newcount\ftno \global\ftno=0
\def\foot{\global\advance\ftno by1\footnote{$^{\the\ftno}$}}
%
\newwrite\ftfile
\def\footend{\def\foot{\global\advance\ftno by1\chardef\wfile=\ftfile
$^{\the\ftno}$\ifnum\ftno=1\immediate\openout\ftfile=foots.tmp\fi%
\immediate\write\ftfile{\noexpand\smallskip%
\noexpand\item{f\the\ftno:\ }\pctsign}\findarg}%
\def\footatend{\vfill\eject\immediate\closeout\ftfile{\parindent=20pt
\centerline{\bf Footnotes}\nobreak\bigskip\input foots.tmp }}}
\def\footatend{}
%
%
\global\newcount\refno \global\refno=1
\newwrite\rfile
\def\ref{[\the\refno]\nref}
\def\nref#1{\xdef#1{[\the\refno]}\writedef{#1\leftbracket#1}%
\ifnum\refno=1\immediate\openout\rfile=refs.tmp\fi
\global\advance\refno by1\chardef\wfile=\rfile\immediate
\write\rfile{\noexpand\item{#1\ }\reflabeL{#1\hskip.31in}\pctsign}\findarg}
\def\findarg#1#{\begingroup\obeylines\newlinechar=`\^^M\pass@rg}
{\obeylines\gdef\pass@rg#1{\writ@line\relax #1^^M\hbox{}^^M}%
\gdef\writ@line#1^^M{\expandafter\toks0\expandafter{\striprel@x #1}%
\edef\next{\the\toks0}\ifx\next\em@rk\let\next=\endgroup\else\ifx\next\empty%
\else\immediate\write\wfile{\the\toks0}\fi\let\next=\writ@line\fi\next\relax}}
\def\striprel@x#1{} \def\em@rk{\hbox{}}
\def\lref{\begingroup\obeylines\lr@f}
\def\lr@f#1#2{\gdef#1{\ref#1{#2}}\endgroup\unskip}
\def\semi{;\hfil\break}
\def\addref#1{\immediate\write\rfile{\noexpand\item{}#1}} 
\def\startrefs#1{\immediate\openout\rfile=refs.tmp\refno=#1}
\def\xref{\expandafter\xr@f}\def\xr@f[#1]{#1}
\def\refs#1{\count255=1[\r@fs #1{\hbox{}}]}
\def\r@fs#1{\ifx\und@fined#1\message{reflabel \string#1 is undefined.}%
\nref#1{need to supply reference \string#1.}\fi%
\vphantom{\hphantom{#1}}\edef\next{#1}\ifx\next\em@rk\def\next{}%
\else\ifx\next#1\ifodd\count255\relax\xref#1\count255=0\fi%
\else#1\count255=1\fi\let\next=\r@fs\fi\next}
%

%
\newwrite\ffile\global\newcount\figno \global\figno=1
\def\fig{fig.~\the\figno\nfig}
\def\nfig#1{\xdef#1{fig.~\the\figno}%
\writedef{#1\leftbracket fig.\noexpand~\the\figno}%
\ifnum\figno=1\immediate\openout\ffile=figs.tmp\fi\chardef\wfile=\ffile%
\immediate\write\ffile{\noexpand\medskip\noexpand\item{Fig.\ \the\figno. }
\reflabeL{#1\hskip.55in}\pctsign}\global\advance\figno by1\findarg}
\def\xfig{\expandafter\xf@g}\def\xf@g fig.\penalty\@M\ {}
\def\figs#1{figs.~\f@gs #1{\hbox{}}}
\def\f@gs#1{\edef\next{#1}\ifx\next\em@rk\def\next{}\else
\ifx\next#1\xfig #1\else#1\fi\let\next=\f@gs\fi\next}
\newwrite\lfile
{\escapechar-1\xdef\pctsign{\string\%}\xdef\leftbracket{\string\{}
\xdef\rightbracket{\string\}}\xdef\numbersign{\string\#}}

\def\writestop{\def\writestoppt{\immediate\write\lfile{\string\pageno%
\the\pageno\string\startrefs\leftbracket\the\refno\rightbracket%
\string\def\string\secsym\leftbracket\secsym\rightbracket%
\string\secno\the\secno\string\meqno\the\meqno}\immediate\closeout\lfile}}
\def\writestoppt{}\def\writedef#1{}
\def\seclab#1{\xdef #1{\the\secno}\writedef{#1\leftbracket#1}\wrlabeL{#1=#1}}
\def\subseclab#1{\xdef #1{\secsym\the\subsecno}%
\writedef{#1\leftbracket#1}\wrlabeL{#1=#1}}
\newwrite\tfile \def\writetoca#1{}
\def\leaderfill{\leaders\hbox to 1em{\hss.\hss}\hfill}
\def\writetoc{\immediate\openout\tfile=toc.tmp
   \def\writetoca##1{{\edef\next{\write\tfile{\noindent ##1
   \string\leaderfill {\noexpand\number\pageno} \par}}\next}}}

%
\catcode`\@=12 
%
\edef\tfontsize{\ifx\answ\bigans scaled\magstep3\else scaled\magstep4\fi}
 \tfontsize  \tfontsize
 \tfontsize \font\titlei=cmmi10 \tfontsize
\font\titleis=cmmi7 \tfontsize \font\titleiss=cmmi5 \tfontsize
\font\titlesy=cmsy10 \tfontsize \font\titlesys=cmsy7 \tfontsize
\font\titlesyss=cmsy5 \tfontsize  \tfontsize
\skewchar\titlei='177 \skewchar\titleis='177 \skewchar\titleiss='177
\skewchar\titlesy='60 \skewchar\titlesys='60 \skewchar\titlesyss='60
 \ifx\answ\bigans\else scaled\magstep1\fi
\ifx\answ\bigans\def\abstractfont{\tenpoint}\else
\font\abssl=cmsl10 scaled \magstep1
\font\absrm=cmr10 scaled\magstep1 \font\absrms=cmr7 scaled\magstep1
\font\absrmss=cmr5 scaled\magstep1 \font\absi=cmmi10 scaled\magstep1
\font\absis=cmmi7 scaled\magstep1 \font\absiss=cmmi5 scaled\magstep1
\font\abssy=cmsy10 scaled\magstep1 \font\abssys=cmsy7 scaled\magstep1
\font\abssyss=cmsy5 scaled\magstep1 \font\absbf=cmbx10 scaled\magstep1
\skewchar\absi='177 \skewchar\absis='177 \skewchar\absiss='177
\skewchar\abssy='60 \skewchar\abssys='60 \skewchar\abssyss='60
\def\abstractfont{\def\rm{\fam0\absrm}
\textfont0=\absrm \scriptfont0=\absrms \scriptscriptfont0=\absrmss
\textfont1=\absi \scriptfont1=\absis \scriptscriptfont1=\absiss
\textfont2=\abssy \scriptfont2=\abssys \scriptscriptfont2=\abssyss
\textfont\itfam=\bigit \def\it{\fam\itfam\bigit}\def\footnotefont{\tenpoint}%
\textfont\slfam=\abssl \def\sl{\fam\slfam\abssl}%
\textfont\bffam=\absbf \def\bf{\fam\bffam\absbf}\rm}\fi
\def\tenpoint{\def\rm{\fam0\tenrm}
\textfont0=\tenrm \scriptfont0=\sevenrm \scriptscriptfont0=\fiverm
\textfont1=\teni  \scriptfont1=\seveni  \scriptscriptfont1=\fivei
\textfont2=\tensy \scriptfont2=\sevensy \scriptscriptfont2=\fivesy
\textfont\itfam=\tenit \def\it{\fam\itfam\tenit}\def\footnotefont{\ninepoint}%
\textfont\bffam=\tenbf \def\bf{\fam\bffam\tenbf}\def\sl{\fam\slfam\tensl}\rm}
\font\ninerm=cmr9 \font\sixrm=cmr6 \font\ninei=cmmi9 \font\sixi=cmmi6
\font\ninesy=cmsy9 \font\sixsy=cmsy6 \font\ninebf=cmbx9
\font\nineit=cmti9 \font\ninesl=cmsl9 \skewchar\ninei='177
\skewchar\sixi='177 \skewchar\ninesy='60 \skewchar\sixsy='60
\def\ninepoint{\def\rm{\fam0\ninerm}
\textfont0=\ninerm \scriptfont0=\sixrm \scriptscriptfont0=\fiverm
\textfont1=\ninei \scriptfont1=\sixi \scriptscriptfont1=\fivei
\textfont2=\ninesy \scriptfont2=\sixsy \scriptscriptfont2=\fivesy
\textfont\itfam=\ninei \def\it{\fam\itfam\nineit}\def\sl{\fam\slfam\ninesl}%
\textfont\bffam=\ninebf \def\bf{\fam\bffam\ninebf}\rm}
%
%

\hyphenation{anom-aly anom-alies coun-ter-term coun-ter-terms}
\def\inv{^{\raise.15ex\hbox{${\scriptscriptstyle -}$}\kern-.05em 1}}

\def\Dsl{\,\raise.15ex\hbox{/}\mkern-13.5mu D} 
\def\dsl{\raise.15ex\hbox{/}\kern-.57em\partial}

\font\bigit=cmti10 scaled \magstep1
\def\lspace{\ifx\answ\bigans{}\else\qquad\fi}
\def\lbspace{\ifx\answ\bigans{}\else\hskip-.2in\fi} 
\def\boxeqn#1{\vcenter{\vbox{\hrule\hbox{\vrule\kern3pt\vbox{\kern3pt
        \hbox{${\displaystyle #1}$}\kern3pt}\kern3pt\vrule}\hrule}}}
\def\mbox#1#2{\vcenter{\hrule \hbox{\vrule height#2in
                \kern#1in \vrule} \hrule}}  
%

\def\darr#1{\raise1.5ex\hbox{$\leftrightarrow$}\mkern-16.5mu #1}

\def\half{{\textstyle{1\over2}}} 
\def\roughly#1{\raise.3ex\hbox{$#1$\kern-.75em\lower1ex\hbox{$\sim$}}}

\overfullrule=0pt
\def\np{Nucl. Phys.}
\def\pl{Phys. Lett.}

\def\cmp{ Comm. Math. Phys.}


\def\Iv{I_{\hat v}}
\def\Lv{L_{\hat v}}

\def\Lv#1{{\cal L}_{{\hat v}_{#1}}}

\def\lva{{\cal L}_{{\hat v}_a}}
\def\lvb{{\cal L}_{{\hat v}_b}}


\def\abeta{\beta_{\mu \nu}}
\def\gmunu{g_{\mu\nu}}
\def\hatg{{\hat g}}
\def\ghat{{\hat g}^{\mu\nu}}
\def\ino{{\psi}_{\mu\nu}}
\def\inoh{{\hat \psi}^{\mu\nu}}
\def\hino{{\hat\psi}}

\def\lc{{\cal L}_c}

\def\lgamma{{\cal L}_{\gamma}}
\def\ab{b_{\mu \nu}}
\def\ci{{\cal C}^i}


\def\ehat{{\hat\eta}^{\mu\nu}}
\def\ehatd{{\hat{\eta}}_{\mu\nu}}
\def\emunu{\eta_{\mu\nu}}
\def\emunua{\eta^{\mu\nu}_a}

\def\het#1{{\hat\eta}_{#1}}
\def\ehatb{{\hat\eta}^{\mu\nu}_b}
\def\ehata{{\hat\eta}^{\mu\nu}_a}
\def\ehato{{\hat\eta}^{\mu\nu}_0}


\def\mgn{{\cal M}_{g,n}}

\def\Ua{{\cal U}_a}

\def\Cel#1{{\cal C}_{#1}}
\def\Cq{{\cal C}_{a_0a_1\ldots a_q}}
\def\x0{{x_0}}
\def\xa{{x_a}}
\def\If#1{{\lambda}_{#1}}
\def\bIf#1{{\bar\lambda}_{#1}}

\def\If#1{{\lambda}_{#1}}
\def\bIf#1{{\bar\lambda}_{#1}}

\def\bel#1{\mu_{#1}}
\def\bbel#1{{\bar\mu}_{#1}}


\def\mgn{{\cal M}_{g,n}}

\def\Ua{{\cal U}_a}

\def\Cel#1{{\cal C}_{#1}}
\def\Cq{{\cal C}_{a_0a_1\ldots a_q}}

\def\za{z_a}
\def\zabar{{\bar z}_a}

\def\Ma{{\rm M}_a}
\def\Mai{{\rm M}_a^{-1}}

\def\Mat{\tilde{\rm M}_a}
\def\Mati{{\tilde{\rm M}}^{- 1}_a}


\def\LPW{J. Labastida, M. Pernici and E. Witten, \np\ {\bf B310} (1988)
611.}
\def\MoSo{D. Montano and J. Sonnenschein, \np\ {\bf B313} (1989) 258.}
\def\MYPE{R. Myers and V. Periwal, \np\ {\bf B333} (1990) 536.}
\def\VV{E. Verlinde and H. Verlinde, \np\ {\bf B348} (1991) 457.}
\def\MYE{R. Myers, \np\ {\bf B343} (1990) 705.}
\def\BMS{R. Brooks, D. Montano and J. Sonnenschein, \pl\ {\bf B214}
(1988) 91.}
\def\MS{D. Montano and J. Sonnenschein, \np\ {\bf B324} (1989) 348.}
\def\BS{L. Baulieu and I.M. Singer, \cmp\ {\bf 135} (1991) 253.}
\def\BCI{C.M. Becchi, R. Collina and C. Imbimbo, hep-th/9311097,
\pl\ {\bf B322} (1994) 79.}
\def\TORINO{C.M. Becchi, R. Collina and C. Imbimbo, hep-th/9406096,
CERN and Genoa preprints CERN-TH 7302/94, GEF-Th 6/1994,
{\it Symmetry and Simplicity in Theoretical Physicis},
Proceedings of the Symposium for the 65-th Birthday of Sergio Fubini,
Turin, 1994 (World Scientific, Singapore, 1994).}

\def\BOTT{R. Bott and L.W. Tu, {\it Differential Forms in Algebraic
Topology} (Springer-Verlag, New York, 1982).}

\def\STORA{R. Stora, F. Thullier and J.C. Wallet, preprint 1994,
ENSLAPP-A-481/94, IPNO-TH-94/29\semi R. Stora, Report at this
Conference.}
\def\BB{L. Baulieu and M. Bellon, \pl\ {\bf B202} (1988) 67.}
\def\GRIB{V. N. Gribov, {\it Instability of non-abelian gauge theories and
impossibiity of choice of Coulomb gauge}, SLAC Translation 176, (1977).}
\def\BI{ C. Becchi and C. Imbimbo, hep-th/9510003,
CERN and Genoa preprints CERN-TH 242/95, GEF-Th 8/1995.}
\def\SIN{I. M. Singer, Commun. Math. Phys. {\bf 60} (1978) 7.}

\def\FUJI{ K. Fujikawa, \np\ {\bf B223} (1983) 218.}

{\nopagenumbers
\null
\vskip -2truecm
\abstractfont\hsize=\hstitle\rightline{
\vtop{
\hbox{hep-th/9511156}
\hbox{GEF-TH/95-13}}}
\pageno=0

\vskip 1truecm
\centerline{\bf A LAGRANGIAN FORMULATION OF }
\bigskip
\centerline{\bf 2-DIMENSIONAL TOPOLOGICAL  GRAVITY}
\bigskip
\centerline{\bf  AND \v{C}ECH-DE RHAM COHOMOLOGY\foot{Report
presented by C. B. at the International Symposium on BRS Symmetry on the
Occasion of its 20th Anniversary, RIMS-Kyoto University,
September 18-22, 1995.}}

\bigskip
\centerline{Carlo Becchi\foot{E-mail: becchi@genova.infn.it},}

\vskip 4pt
\centerline{\it Dipartimento di Fisica dell' Universit\'a di Genova}

\centerline{\it Via Dodecaneso 33, I-16146, Genova, Italy}
\vskip 7pt
\centerline{Camillo Imbimbo\foot{E-mail: imbimbo@vxcern.cern.ch}}
\vskip 4pt
\centerline{{\it Theory Division, CERN, CH-1211 Geneva 23,
Switzerland}\foot{On leave from INFN, Sezione di Genova, Genoa, Italy}}
\ \medskip
\centerline{ABSTRACT}
We present a very simplified analysis of how one can overcome the
Gribov problem in a non-abelian gauge theory. Our formulae, albeit
quite simplified, show that possible breakdowns of the Slavnov-Taylor
identity could in principle come from singularities in space of gauge
orbits.  To test these ideas we exhibit the calculation of a very
simple correlation function of 2-dimensional topological gravity and
we show how in this model the singularities of the moduli space
induce a breakdown of the Slavnov-Taylor identity.  We comment on the
technical relevance of the possibility of including the singularities
into a finite number of cells of the moduli space.

\ \vfill
\ifx\answ\bigans
\leftline{October 1995}
\else\leftline{October1995}\fi
\eject}
\ftno=0

\newsec{Introduction}

The aim of this report is to discuss to what extent a functional integral
approach to 2-dimensional topological gravity \ref\bi{\BI}
 is relevant to the study
of gauge theories in the presence of Gribov horizons \ref\grib{\GRIB}.
 This is the
situation in which the Faddeev-Popov determinant is not positive definite
and the corresponding measure becomes singular.

Due to the Gribov phenomenon the Feynman functional integral of a gauge
theory cannot be defined as a unique integral over a single connected
domain. In a general situation \ref\sin{\SIN},
 if the gauge freedom has been fixed
according to the Landau background gauge prescription, one has to choose
a suitable set of background fields $\{V^i\}$ and a corresponding set of
cells, whose characteristic functions are
$\{\chi_i (A)\}$, covering the whole field functional space and such that
for some suitable, positive, $\{a^i\}$ :
$$\chi_i (A)\equiv\chi_i (A)\,\,
\Theta\bigl[a^i-\int d^4x\sum_{\mu}\bigl(V^i_{\mu}-A_{\mu}\bigr)^2
\bigr]\ ,$$
where $\Theta$ is the Heavyside function.

In the $i^{th}$ cell the gauge fixing condition is:
$$\nabla_{V^i,\,\mu}\bigr(A-V^i\bigr)_{\mu}=0\ ,$$
where the sum over the repeated indices is understood and:
$$\nabla_{V,\,\mu}\equiv\partial_{\mu}-g\, V^i_{\mu}\wedge\ ,$$
is the covariant derivative corresponding to the connection $V$ and the
wedge product $\wedge$ is defined by the gauge group structure
constants.

Then the question comes on how should one define the Feynman functional
$Z$. Naively, for any cell, one would construct the action $S_i$
 given by:
$$ S_i\equiv S^{(inv)}+s\int d^4x\,\bigl( \bar c\cdot
\nabla_{V^i,\,\mu}\bigr(A-V^i\bigr)_{\mu}\bigr)\ ,$$
where $S^{(inv)}$ is gauge invariant and independent of $V^i$ and $s$ is
the usual BRS operator acting only on the quantized fields, and
one would write:
$$Z\equiv\sum_i\int d[\Phi]\,\chi_i\, {\rm e}^{-S_i}\ ,$$
However in general this expression is sick since, for two
neighbouring cells $i$ and $j$:
$${\rm e}^{-S_i}\not={\rm e}^{-S_j}\ ,$$
and therefore $Z$ is affected by small deformations of the boundaries
of the cells.
Indeed, introducing the interpolating action:
$$S_{ij}\equiv t\, S_i+(1-t)\, S_j\ ,$$
 one has:
$${\rm e}^{-S_j}-{\rm e}^{-S_i}=s\bigl[\int d^4x
\bigl(V^i-V^j\bigr)_{\mu}(x)\cdot\nabla_{A,\,\mu}\bar c (x)\int_0^1 dt\,\,
{\rm e}^{-S_{ij}(t)}\bigr]\ . \quad\quad \quad\quad(1)$$

It is also apparent that the naive definition of the functional integral
 violates the Slavnov-Taylor identity. Let us remember that this identity
is the faithful translation of the gauge invariance of the
theory in terms of the Green functional generator. It asserts the vanishing
of the expectation value of any s-exact operator, that is of any operator
of the form $sX$. This vanishing follows from the possibility of
performing functional integrations by parts disregarding boundary
contributions.
This is however not possible in the present situation.

If the whole functional space were covered by only two cells
( 1 and 2 ), and hence: $$\chi_1+\chi_2=1\ ,$$ Eq.(1)
would suggest the alternative definition:
$$\eqalign{& Z\equiv
 \sum_{i=1}^2\int d[\Phi]\,\chi_i\, {\rm e}^{-S_i}\,\cr & -
\int d[\Phi]\,s\chi_1\int d^4x
\bigl(V^1-V^2\bigr)_{\mu}(x)\cdot\nabla_{A,\,\mu}\bar c (x)\int_0^1
dt\,\,
{\rm e}^{-
S_{12}(t)}\quad\quad\quad\quad\quad\quad\quad\quad(2)\cr&=
 \sum_{i=1}^2\int d[\Phi]\,\chi_i\, {\rm e}^{-S_i}\cr & -
\int d[\Phi]\int d^4y\,\nabla_{A,\,\nu} c (y)\cdot
{\delta\,\chi_1\over\delta A_{\nu}(y)}\int d^4x\,
\bigl(V^1-V^2\bigr)_{\mu}(x)\cdot\nabla_{A,\,\mu}\bar c (x)\int_0^1 dt
{\rm e}^{-S_{12}(t)}\,\ ,\cr}$$
that is not affected anymore by the above mentioned sickness.

It is rather clear that the second term in the right-hand
side of this formula does not vanish except for very special choices of
the observables that appear in the invariant part of the action. Therefore
the last term in the right-hand side of Eq.(2) gives a relevant
contribution to the definition of the Feynman vacuum functional.

It is also easy to see that the Slavnov-Taylor identity is recovered
by the introduction of the boundary term. Indeed owing to Eq.(2) we get:
$$\eqalign{& <sX>={1\over Z}\bigl[\sum_{i=1}^2\int d[\Phi]\,\chi_i\,
{\rm e}^{-S_i}\, sX
 \,\cr & -
\int d[\Phi]\,s\chi_1\int d^4x
\bigl(V^1-V^2\bigr)_{\mu}(x)\cdot\nabla_{A,\,\mu}\bar c (x)\int_0^1
dt\,\,
{\rm e}^{-
S_{12}(t)}\, sX\bigr]\ ,\cr}$$
We can compute the numerator in the right-hand side of this equation
using the identity: $s\chi_1+s\chi_2=0$ and integrating by parts, getting:
$$\eqalign{&Z<sX>=-\sum _{i=1}^2\int d[\Phi]\,s\chi_i\,
{\rm e}^{-S_i}\, X\cr & -
\int d[\Phi]\,s\chi_1\, s\bigl[\int d^4x
\bigl(V^i-V^j\bigr)_{\mu}(x)\cdot\nabla_{A,\,\mu}\bar c (x)\int_0^1 dt\,\,
{\rm e}^{-S_{ij}(t)}\bigr]\,X\cr & =\int d[\Phi]\,s\chi_1\bigl[
{\rm e}^{-S_2}-{\rm e}^{-S_1}\bigr]\, X\cr & -
\int d[\Phi]\,s\chi_1\, s\bigl[\int d^4x
\bigl(V^i-V^j\bigr)_{\mu}(x)\cdot\nabla_{A,\,\mu}\bar c (x)\int_0^1 dt\,\,
{\rm e}^{-S_{ij}(t)}\bigr]\,X\, =0\ .\cr }$$

The natural question is now how Eq.(2) generalises to the case of a more
complex covering, an answer to this question is known in the case
in which the covering involves only a finite number of cells. Indeed in
this case the natural generalisation of Eq.(2) is given by the
construction of a \v{C}ech-de Rham cocycle. In the case in which the
covering requires an infinite number of cells to our knowledge
the generalisation of Eq.(2) in not known. We imagine that in the general
framework involving local observables the number of cells is infinite and
hence the whole construction becomes singular.

The next sections of this report
present in some details the application of this construction to the
case of two dimensional topological gravity. The situation in this
model is different from the gauge theory case, since, strictly speaking,
in the topological case any field configuration lies on the Gribov horizon
that has carefully been avoided in the gauge case. However, by means of a
second
Faddeev-Popov procedure that leads to the dynamical interpretation of the
moduli of the Riemann surface playing the role of the world-sheet of the
model,
one is led to a problem that, although finite dimensional, presents
interesting
analogies with the gauge case. In particular the configuration space, that is
the moduli space, is topologically non-trivial. The Feynman integral over it
can
be computed using a finite decomposition in cells  that however contain
singular
 points of the moduli space. We shall see that these singular points are
 responsible for the non-trivial properties of the model that originate from
a breakdown of the Slavnov-Taylor identity.

The presence of worse singularities in the infinite dimensional situation
leaves open the possibility of analogous phenomena.

\newsec{Two-dimensional topological gravity}

Two dimensional topological gravity \ref\lpw{\LPW},\ref\moso{\MoSo}\
turns out to be a particularly
interesting  toy model to study the role of Gribov horizons and in
particular of the singularities of the moduli space, which is the
space of the physically relevant configurations.

We shall see in a moment that in our model the Faddeev-Popov
determinant vanishes all over the gauge orbit space, and hence
the Gribov horizon coincides with the space of moduli. Therefore
a second Faddeev-Popov procedure is needed in order to define a
non-degenerate functional measure and for this the action of BRS
operator has to be extended to the global quantum mechanical
variables that coincide with the moduli and with their
supersymmetric partners, the supermoduli.

The new Faddeev-Popov measure, including both the local fields and
the global quantum mechanical variables, is generically non-degenerate.
The Gribov horizon associated with it has codimension one on moduli space.
This implies that the functional integral defines correlators of
observables which are local closed top-forms on the moduli space.
In order that the correlators have physical meaning, however, such
locally defined forms must be local restrictions of forms which are
globally defined.

 In our case, functional averages  of BRS closed operators are not in
general globally defined. We will demonstrate this by deriving the Ward
identities associated to finite reparametrizations of the background
gauge. This phenomenon is originated by the dependence of the
observables on derivatives of the super-ghost field.

We shall see that even if functional averages of BRS closed operators are
not globally defined, it is still possible to associate to them
globally defined forms by resorting to the \v{C}ech-De Rham
 cohomology \ref\bott{\BOTT} if  the operators
 correspond  to elements of the equivariant BRS
cohomology \ref\stora{\STORA}.

In a previuos paper we have derived
 from our ``anomalous'' Ward identities a chain of descendant
identities defining a local cocycle of the \v{C}ech-De Rham complex of the
moduli space. A well-known construction of cohomology theory leads from
this local cocycle to a globally defined form. The integral of the globally
defined form receives contributions not only from the original local top-
form (which vanishes in the superconformal gauge), but also from the
tower of local forms of lower degree that solve the chain of Ward identities.

A very important point that emerges from our analysis is that in the great
majority of situations the correlators, when computed in the superconformal
gauge, receive
non-trivial contributions only from the singularities of moduli space.
Indeed, these singularities, which correspond to nodes
of the punctured Riemann surface that is identified with the
world-sheet of our model, lie generally at the interior of Gribov domains.
The correlators turn out to vanish, and hence to be locally integrable, in the
neighbourhoods of the singularities of moduli space. These singularities,
however, determine the lack of global definition of the local correlators
which in fact turn out to correspond to non-trivial elements
of \v{C}ech-De Rham cohomology.

We now come to the description of the model. Being a field theory model it
involves a set of local variables that are identified with the two-dimensional
metric $\gmunu$, the gravitino field $\ino$, the ghost field
$c^{\mu}$  associated with the local gauge degrees of freedom, that is with
diffeomorphisms, and its superpartner $\gamma^{\mu}$.
The theory is defined by its invariance under the nilpotent BRS
transformations \ref\bms{\BMS},\ref\mype{\MYPE},\ref\vv{\VV}:
$$\eqalign{& s\, \gmunu = \lc \gmunu + \ino \cr
& s\, \ino = \lc \ino - \lgamma \gmunu \cr}\qquad
\eqalign{&s\, c^{\mu}\, = \half \lc c^{\mu} + \gamma^{\mu} \cr
&s\, \gamma^{\mu}=\lc \gamma^{\mu}\cr}$$

Notice that the presence of the gravitino and of the superghost field makes
the cohomology of the $s$-operator trivial and hence determines the
topological nature of the model.

It is convenient to decompose the metric into the corresponding reduced
metric (complex structure) and the Liouville field according:
$$\gmunu (x) \equiv \sqrt{g}
\gmunu(x) \equiv e^{{\varphi}}
\hatg_{\mu\nu} (x)$$
with:
$${\rm det}({\hat g})_{\mu\nu}=1$$
We also introduce the traceless gravitino field:
$$\inoh \equiv \sqrt{g}(\psi^{\mu\nu} -\half
g^{\mu\nu}\psi^{\sigma}_{\sigma})
$$
The conformal background gauge prescription corresponds to the
conditions:
$$\gmunu (x) = \eta_{\mu\nu} (x;m)\ \longrightarrow\
\hatg_{\mu\nu}=\ehatd ,\qquad\varphi=\bar\varphi$$
$$\inoh = \hat f^{\mu\nu} $$
that are implemented under the functional integral by a system of
Lagrange multipliers and antighost fields, whose
BRS transformations are given by:
$$\eqalign{
&s\,\ab = \Lambda_{\mu \nu}\cr
&s\,\abeta = L_{\mu \nu}\cr
&s\,\chi = \lc \chi + \pi\cr }\qquad
\eqalign{&s\,\Lambda_{\mu \nu} = 0\cr
&s\,L_{\mu \nu} = 0 \cr
&s\,\pi = \lc \pi -\lgamma \chi\cr}$$
Due to the above mentioned triviality of the $s$-cohomology, the
Lagrangian of the model coincides with the gauge fixing term
\vv,\ref\torino{\TORINO}:
$$
\eqalign{{\cal L}  = s \bigl[\half  b_{\mu \nu} (\ghat -\ehat)  +
\half \beta_{\mu \nu}
(\inoh - \hat f^{\mu\nu} )  + \chi \partial_{\mu}(\ghat
\partial_{\nu}(\varphi -{\bar \varphi}))  \bigr]}$$
This completes the local description of our theory.

Concerning its global
properties, we identify the world-sheet with a punctured Riemann surface
and we limit our study to the observables that correspond to local
operators sitting on fixed points of the surface. It is obvious that, the
theory being a gauge theory, these operators have to be $s$-closed.

It is well known that if the surface has genus $g$ and contains $n$ fixed
points ( we take $3g-3+n>0$ ) our Lagrangian is degenerate since the field
$b$ has $3g-3+n>0$ zero-modes. Indeed this is the number of independent
deformations of the background metric that do not correspond to
coordinate transformations. Let :
$$\int d^2x b_{\mu\nu}{\hat \eta}_i^{\mu\nu}\quad{\rm for}\quad
i=1,\ldots,3g-3+n$$
define these zero modes, the ${\hat \eta}_i^{\mu\nu}$'s are interpreted as
derivatives of the background complex structure with respect to the
moduli of our theory. Fixing these zero modes automatically introduces
the moduli into the dynamics of our theory.

We also introduce a set of Grassmann variables $\{p^i\}$, that we call
supermoduli and that we assume to transform as 1-forms under moduli space
reparametrizations; we choose the gravitino background as follows
$$\hat f^{\mu\nu}= d_p\,\ehat \qquad  d_p \equiv p^i {\partial \over \partial
m^i}\ .$$

In order to fix the $b$-field zero modes we have to introduce a further set
of Lagrange multipliers that are not local field but global variables.
Then, to keep the original $s$-exact structure of the Lagrangian, we
extend the definition of the coboundary operator $s$ on these new
parameters \ref\bb{\BB},\torino:
$$\eqalign{
s\, m^i = &\, \ci \cr
s\, p^i = & -\Gamma^i\cr}\qquad
\eqalign{s\,\ci =& 0 \cr s\,\Gamma^i = &0 \cr}$$
With the new definition of the coboundary operator the Lagrangian
becomes \torino:
\eqn\lagr{\eqalign{
{\cal L} =& \half \Lambda_{\mu \nu}(\ghat -\ehat) + \half L_{\mu \nu}
(\inoh  - d_p\ehat )  - \half \ab \lc \ghat  - \half \abeta \lgamma \ghat
\cr & +\half \inoh \left[ (\lc\beta)_{\mu \nu} + \ab + 2\partial_{\mu}\chi
\partial_{\nu}(\varphi -{\bar \varphi})\right]+\pi \partial_{\mu} (\ghat
\partial_{\nu} (\varphi -{\bar
\varphi}))
- \chi \partial_{\mu} (\ghat \partial_{\nu}\psi^{\prime})\cr
&+\half\abeta d_{\Gamma}\ehat + \half\ab d_C \ehat  +
\half\abeta d_p d_C \ehat + \chi \partial_{\mu}(\ghat
\partial_{\nu}d_C{\bar\varphi})\ ,\cr}}
with:
$$\psi^{\prime} \equiv {\bar D}_{\sigma} c^{\sigma} +
\half\psi^{\sigma}_{\sigma}.$$
It is apparent that the eighth term in the right-hand side of
\lagr\ fixes the zero modes of the $b$  field.

Having so specified a non-degenerate action for our model we come to the
selection of observables. It is clear that the usual
prescription that the observables be elements of the $s$-cohomology does
not work here since this cohomology is empty; we have therefore to
enlarge our study and consider $s$-exact operators.
Furthermore we consider a natural assumption that observables be
independent of the Lagrange multipliers. That is:

\eqn\uno{{\delta \Omega\over\delta \Lambda_{\mu\nu}}=
{\delta \Omega\over\delta L_{\mu\nu}}=0\ .}
In the standard situation this hypothesis is not necessary since \uno\
is automatically verified within the $s$-cohomology elements, here \uno\ is
assumed since otherwise the observables would affect the gauge fixing.
We assume furthermore that our observable algebra be generated by local
operators sitting on the punctures of the world-sheet. Therefore a generic
element of it is a linear combination of operators of the form:
\eqn\nome{\Omega=\prod_k O_k (P_{i_k})\ .}
Under the above assumptions one can directly substitute $
\ghat \rightarrow \ehat $ and $\inoh \rightarrow d_p \ehat $
 into the Lagrangian, that becomes:
\eqn\due{\eqalign{
{\cal L}^{\prime} = &\half\bigl[ -\ab \lc \ehat - \abeta \lgamma \ehat
+d_p\ehat (\lc \beta)_{\mu\nu}\cr
& +\ab (d_C \ehat - d_p \ehat ) + \abeta d_{\Gamma}\ehat +
\abeta d_p d_C \ehat \bigr] .\cr}}
If, furthermore, the observables are independent of
antighost fields (a sufficient, weaker hypothesis would exclude only the
antighost field zero modes):
$${\delta \Omega\over\delta b_{\mu\nu}}  =
{\delta \Omega\over\delta \beta_{\mu\nu}}=0\ ,$$
the last term in the right-hand side of \due\ does not contribute to
the Feynman functional integral.

Let us now consider the Feynman integral of our model.
The correlation functions should be computed by performing
the functional integration over the local fields
$ g, \psi , c, \gamma , b, \beta ,  L, \Lambda $,
that we collectively label by $\Phi$. However, the
presence of zero modes requires that we also integrate over the Lagrange
multipliers ${\cal C} $ and $ \Gamma$. With this definition of the Feynman
integral the correlators are functions of the moduli $m$ and of the
supermoduli $p$.
Remembering that the coboundary operator does not leave invariant these
variables it is natural to include them among the set of the integration
variables; for a moment, however, it is interesting to refrain from
doing this.

It is a natural question whether the Slavnov-Taylor identities, that in
the standard situation insure the vanishing of the expectation values of
$s$-exact operators, hold true in our case.
We therefore consider  a generic $b,\, \beta ,\, \Lambda ,\, L  ,\, m,\, p,\,
{\cal C}, $ and $\Gamma $ independent operator that we call {\it
admissible} and we notice first of all that, if $\Omega$ is admissible:
\eqn\tre{\int\,[d\Phi] e^{-S(\Phi ;m^i,\,p^i)}\Omega =\prod_i \delta (\ci -
p^i)
\prod_j\delta (\Gamma^j)\langle\Omega \rangle_{m,p}\ .}
Indeed the $\delta$ functions are generated by the integral over the $b$
and $\beta$ zero modes.

Now we come to the Slavnov-Taylor identities considering a $s$-exact
admissible operator $sX$ and we get:
\eqn\st{\eqalign{&\langle s\, X(\Phi)\rangle_{m,p} \equiv
\int\prod_i d\ci d\Gamma^i \int\,[d\Phi] e^{-S(\Phi ;m^i,\,p^i)}sX
\cr &=\int\prod_i d\ci d\Gamma^i \int\,[d\Phi] \bigl(s_L+\ci\partial_{m^i}-
\Gamma^i\partial_{p^i}\bigr)e^{-S(\Phi ;m^i,\,p^i)}X
\cr &=\int\prod_i d\ci d\Gamma^i \bigl(\ci\partial_{m^i}-
\Gamma^i\partial_{p^i}\bigr)\int\,[d\Phi] e^{-S(\Phi ;m^i,\,p^i)}X
\cr &=\int\prod_i d\ci d\Gamma^i p^i\partial_{m^i}
\int\,[d\Phi] e^{-S(\Phi ;m^i,\,p^i)}X
= d_p \langle X(\Phi)\rangle_{m,p} \ ,\cr}}
where we have introduced $s_L$, i.e. the restrictions of the coboundary
operator $s$ to the local fields, and we have used the fact that its
integral over $\Phi$ vanishes after integration by parts.
We have also used \tre\ for $X$.

This equation shows that the Slavnov-Taylor identity is violated by a
locally exact form on the moduli space. If the breaking were a globally
exact form, by integrating the correlators over the moduli superspace,
that is, by integrating over the moduli space the top form corresponding
to the coefficient of the term of degree $6g-6+2n$ in $p$, one would  recover
the unbroken Slavnov-Taylor identity.
This, however, would imply the
triviality of our construction since, due to the triviality of $s$-cohomology,
all the correlators of our model would vanish.

It is therefore essential to understand if the above form is globally defined.
For this we have to examine the structure of the moduli space $\mgn$.

The background metric $\emunu (x;m)$ cannot be chosen to be a
everywhere
continuous function of the moduli space. In fact $\emunu (x;m)$ is a
section of
the gauge bundle over $\mgn$ defined by the space of two-dimensional
metrics on a surface of given genus and $n$ punctures. This bundle
is non-trivial and therefore does not admit a global section.
It follows that $\emunu (x;m)$ must be a {\it local} section of
the bundle of two-dimensional metrics. Let $\{\Ua\}$ be a covering of
the moduli space. The background gauge is defined by a collections
$\{\emunua (x;m)\}$ of two-dimensional metrics, with each $\emunua
(x;m)$
 defined, as a function of $m$, on $\Ua$.

Let $\ehata$ and $\ehato$ be two gauge-equivalent reduced
metrics (complex structures):
\eqn\metric{\ehata (\xa;m)=
{1\over {\rm det}({\partial\xa\over \partial\x0})}
{\partial \xa^{\mu}\over \partial \x0^{\sigma}}
{\partial \xa^{\nu}\over \partial \x0^{\rho}}
{\hat\eta}_0^{\sigma\rho}(\x0;m),}
related by a diffeomorphism which may in general depend on $m$:
\eqn\diffeo{\xa \rightarrow \x0 (\xa;m).}
{}From \metric\ one derives the transformation law for
${{\partial{\hat\eta}^{\mu\nu}_0}\over {\partial m^i}}
= \partial_i{\hat\eta}^{\mu\nu}_0$:
\eqn\dmetric{\partial_i{\hat\eta}^{\mu\nu}_a (\xa;m)=
{1\over {\rm det}({\partial\xa\over \partial\x0})}
{\partial \xa^{\mu}\over \partial \x0^{\sigma}}
{\partial \xa^{\nu}\over \partial \x0^{\rho}}
\bigl( \partial_i {\hat\eta}^{\sigma\rho}_0 (\x0;m) +
({\cal{L}}_{v^i_a}{\hat\eta}_0)^{\sigma\rho}
(\x0;m)\bigr ) , }
where $v_a$ is the vector field defined by the equation
\eqn\vector{ v^{\mu}_{i a} = \partial_i x^{\mu}_0
(\xa;m)\vert_{\xa=\xa(\x0;m)}.}
We also define:
\eqn\hvector{{\hat v}_{\mu}^a\equiv p^i v^{\mu}_{i a}\ .}
The action of our model is invariant under background coordinate
transformations that are independent of the moduli; if the observables
are admissible and transform like scalars under world-sheet coordinate
transformations, the whole effect of the transition from $\ehato$ to
$\ehata$ is reduced to the substitution:
$\partial_i{\hat\eta}^{\mu\nu}_a (\xa;m)\rightarrow
\partial_i {\hat\eta}^{\sigma\rho}_0 (\x0;m) +
({\cal{L}}_{v^i_a}{\hat\eta}_0)^{\sigma\rho}
(\x0;m)\ , $ and $c\rightarrow c+\hat v_a\ .$
This result is a consequence of \tre\ .

Therefore we restrict, from now on, the admissible local observables
to those that transform as scalars under coordinate transformations.
Under this condition we can compute the correlation functions in
the $\ehata$ background in terms of those in the $\ehato$ one.
Labelling by $S_0$ the action corresponding to $\ehato$, we get:

$$\eqalign{&\langle \Omega\rangle_{\het{a}}
\equiv  \int\prod_i d\ci d\Gamma^i \int d[\Phi] e^{- S_a }\Omega (c,\dots)\cr
&\quad = \int\prod_i d\ci d\Gamma^i \int d[\Phi] e^{- S_0 + s\int d^2\, x
\abeta(\lva{\hat{\eta}}_0)^{\mu\nu}}\Omega (c+\hat v_a,\ldots)\  .\cr}$$
We consider now the correlation functions of the
same observable $\Omega$ in two background metrics
$\ehata$ and $\ehatb$ that are gauge-equivalent to $\ehato$. We have:
\eqn\corra{\langle\Omega\rangle_{\het{a}}
=  \int\prod_i d\ci d\Gamma^i \int d[\Phi] e^{- S_0 + s\int d^2\, x
\abeta(\lva{\hat{\eta}}_0)^{\mu\nu}}\Omega (c+{\hat v}_a,\ldots),}
and
\eqn\corrb{\langle \Omega\rangle_{\het{b}}
= \int\prod_i d\ci d\Gamma^i  \int d[\Phi] e^{- S_0 + s\int d^2\, x
\abeta(\lvb{\hat{\eta}}_0)^{\mu\nu}}\Omega(c+{\hat v}_b,\ldots).}
In order to compare the correlation functions \corra\ and \corrb\ we
introduce the interpolating action:
\eqn\actionint{ S_{ab}(t) = \, S_0 -
s \int d^2\, x \abeta \left[ t(\lvb{\hat{\eta}}_0)^{\mu\nu} )
+ (1 - t)(\lva{\hat{\eta}}_0)^{\mu\nu})\right]\ ,}
in terms of which we can compute the difference:
\eqn\difab{\eqalign{&\langle \Omega\rangle_{\het{b}} -
\langle \Omega\rangle_{\het{a}}\equiv \left(\delta\langle
\Omega\rangle\right)_{ab} \cr
&= \int_0^1 dt\int\prod_i d\ci d\Gamma^i \int d[\Phi]{d\over {dt}}
\left[e^{- S_{ab}(t)}\Omega\left(c + t{\hat{v}}_b + (1 - t){\hat{v}}_a,
\gamma, \hatg, \hino\right)\right] \cr
&= -\int_0^1dt\int\prod_i d\ci d\Gamma^i \int d[\Phi]{d S_{ab}(t)\over dt}
e^{- S^{\prime}_{ab}(t)}
\Omega\left(c + t{\hat{v}}_b + (1 - t){\hat{v}}_a,
\gamma, \hatg, \hino\right) \cr
&+ \int_0^1dt\int\prod_i d\ci d\Gamma^i
\int d[\Phi]e^{- S_{ab}(t)}\int d^2 x \left({\hat{v}}^{\mu}_b
- {\hat{v}}^{\mu}_a\right)
{{\partial}\over {\partial c^{\mu}}}
\Omega\left(c + t{\hat{v}}_b + (1 - t){\hat{v}}_a,
\gamma, \hatg, \hino\right). \cr}}

Taking into account the structure of the interpolating action
and \st\ it is clear that the first term in the right-hand side
of this equation corresponds to a locally exact form; this is however
not true for the second term.
Thereby we see that the restriction of the observable algebra to
that generated by the admissible scalar operators is not sufficient
to guarantee that the difference of the local top forms corresponding
to the correlators in two different backgrounds be a locally exact form
on moduli space. Remember that, as noticed at the beginning of this report,
 if this condition is not verified, the correlators depend on the particular
 choice of the particular covering chosen for $\mgn$.

The second term in the right-hand side of \difab\ in general does not
vanish. However, recalling that the local operators in \nome\ are
sitting on fixed points of the Riemann surface and hence the vector
fields ${\hat{v}}$ vanish at these points, we see that the unwanted
term vanishes if the observable does not depend on the derivatives of
the $c$ field. That is, if the observable depends on the $c$ field
only through its value at the punctures.  It is shown in \stora\
that this further restriction reduces the local observables to the
elements of the {\it equivariant cohomology } of $s$.

With this final restriction of the observables we obtain:
\eqn\piff{\left(\delta\langle \Omega\rangle\right)_{ab}
= d_p \int_0^1dt \int\prod_i d\ci d\Gamma^i  \int d[\Phi]
e^{- S_{ab}(t)}\left(I_{\hat{v}_b} - I_{\hat{v}_a}\right)\Omega}
with:
$$\Iv \equiv -\int d^2\,x {\hat{v}}^{\mu}(x) {\delta\over\delta
\gamma^{\mu}(x)}$$
The possibility of generalising this equation to the case of many
overlapping covers is a crucial result of our analysis that allows
a complete characterisation of the correlators as elements of
\v{C}ech-De Rham cohomology \bi.
Indeed, one extends iteratively the
definition of the
difference operator $\delta$ to the case of many intersecting covers
as follows:
\eqn\dimone{
\left(\delta\langle \Omega \rangle \right)_{a_0\ldots a_{q+1}} \equiv
\sum_{l=0}^{q+1} (-1)^l \left(\langle \Omega \rangle \right)_{a_0\ldots
\check{a}_l\ldots a_{q+1}}\ ,}
where the check mark above $a_l$ means that this simbol should be
omitted.
Then, using exactly the same method as above, one finds:
\eqn\cdrh{\bigl(\delta\langle\Omega\rangle\bigr)_{a_0\ldots a_q}=
d_p\bigl(\langle\Omega\rangle\bigr)_{a_0\ldots a_q}.}
In \dimone\ and \cdrh\ one has introduced:
\eqn\boh{\eqalign{\bigl(\langle \Omega\rangle \bigr)_{a_0\ldots a_q}
&\equiv \sum_{k=0}^q (-1)^k\int_0^1 \prod_{l=0}^q dt_l
\delta\Bigl(\sum_{j=0}^q t_j -1\Bigr)\cr
&\qquad\int\prod_i d\ci d\Gamma^i
\int d [\Phi] {\rm e}^{-S_{a_0\ldots a_q}(t_0,\ldots,t_q)}\,
I_{a_0},...,\check{I}_{a_k},...,I_{a_q} \Omega, \cr }}
and the interpolating action is given by:
\eqn\interpolaction{ S_{a_0\ldots a_q}(t_0,\ldots,t_q)=
S_0 - s\,\sum_{k=0}^q t_k \int d^2x \abeta
\bigl(\Lv{a_k}\het{0}\bigr)^{\mu\nu}. }
A detailed discussion of the structure of a correlator as element of the
\v{C}ech-De Rham cohomology can be found in \bi.
For the purposes of the present report it is sufficient to discuss the
simplest non-trivial application of our method.

However it is useful to mention that, given a
\v{C}ech-De Rham cocycle, the corresponding moduli space integral is
computed as follows. Let $\{\Cel{a}\}$
be a cell decomposition of $\mgn$, with $\Cel{a} \subset \Ua$, and let
$\Cq $ of codimension
$q$ in $\mgn$ oriented in such a way that the boundary of a cell
$\partial\Cel{a_0a_1\ldots a_q}$ satisfies:
\eqn\orient{\partial\Cel{a_0a_1\ldots a_q} =\cup_b\Cel{a_0a_1\ldots
a_qb},}
where we have introduced the convention that
$\Cel{a_0a_1\ldots a_q}$ is antisymmetric in its indices in the sense
that it changes orientation when
exchanging a pair of indices. We have defined
in this way
$q$-chains of cells of codimension $q$ that are adjoint to the $q$-cochains
defined by $\left(\langle \Omega \rangle \right)_{a_0,\ldots,a_q} $. Given
a $q$-chain and a
$q$-cochain, we can define the integral:
\eqn\integralchain{\int_{{\cal C}_q}\langle \Omega \rangle \equiv
\sum_{a_0 <a_1
\ldots<a_q} \int_{\Cq} \left(\langle \Omega \rangle \right)_{a_0a_1\ldots
a_q}.}

To study explicitly an example we still have to find a suitable set of
local observables satisfying our constraints, that is corresponding to
non-trivial classes of the $s$-equivariant cohomology.

The wanted set of observables can be constructed
\mype,\ref\mye{\MYE},\ref\ms{\MS},\ref\bs{\BS},\vv,\ref\bci{\BCI}
starting from the Euler two-form
\eqn\euler{\sigma^{(2)} = {1\over 8\pi}\sqrt{g} R \epsilon_{\mu \nu}
dx^{\mu}\wedge dx^{\nu},}
where $R$ is the two-dimensional scalar curvature and $\epsilon_{\mu
\nu}$
is the antisymmetric numerical tensor defined by $\epsilon_{12}=1$.
Since $s$ and the exterior differential $d$ on the two-dimensional
world-sheet anti-commute, the two-form in Eq. \euler\
gives rise
to the descent equations:
\eqn\descent{\eqalign{
s \sigma^{(2)} =& d \sigma^{(1)}\cr
s \sigma^{(1)} =& d \sigma^{(0)} \cr
s \sigma^{(0)} =& 0. \cr}}
The 0-form $\sigma^{(0)}$ and the 1-form $\sigma^{(1)}$
are computed to be
\eqn\observables{\eqalign{
\sigma^{(0)} =& {1\over 4\pi}\sqrt{g} \epsilon_{\mu \nu}
\left[{1\over 2}c^{\mu} c^{\nu}R + c^{\mu}D_{\rho}(\psi^{\nu\rho}
- g^{\nu\rho}\psi^{\sigma}_{\sigma}) + D^{\mu}\gamma^{\nu} -{1\over
4}\psi^{\mu}_{\rho}\psi^{\nu\rho}\right] \cr
\sigma^{(1)} =& {1\over 4\pi}\sqrt{g}\epsilon_{\mu\nu}\left[c^{\nu} R +
D_{\rho}(\psi^{\nu \rho} - g^{\nu\rho}\psi^{\sigma}_{\sigma})
\right]dx^{\mu}.\cr}}

$\sigma^{(0)}$ correspond to non-trivial class in the equivariant
cohomology of $s$,
in particular it is clear that it satisfies our constraints. Therefore we shall
choose in general:
 $$\Omega=\prod_i\bigl(\sigma^{(0)}(P_i)\bigr)^{n_i}$$

\newsec{The study of a very simple physical expectation value}

Now we give some formulae that are useful for explicit calculations.
Identifying $\x0^{\mu}$ with the isothermal coordinate frame, i.e.
${\hat \eta}_0^{\mu\nu}=\delta^{\mu\nu}$, and given a generic choice of the
background metric $\het{a}$, we introduce the transition matrix:
\eqn\transition{ (\Ma)^{\mu}_{\,\,\nu}
\equiv {\partial x_0^{\mu}\over \partial x^{\nu}_a}\ . }
We have:
\eqn\change{(\het{a})^{\mu\nu}=
\det(\Ma)(\Mai\Mati)^{\mu\nu},}
where $\Mat$ is the transposed of $\Ma$.
We also have the vector field ${\hat{v}}_a$ (see \vector\ ) given by:
\eqn\vectorr{ {\hat v}^{\mu}_a = d_p x_0^{\mu} (\xa ;
m)\vert_{\xa=\xa(x_0 ; m)}.}
and satisfying the following identities:
\eqn\devec{{{\partial{\hat v}^{\mu}_a}\over
{\partial x_0^{\lambda}}}
= \left(d_p\Ma \Mai\right)^{\mu}_{\,\,\lambda}, }
and:
\eqn\dmvec{d_p{\hat v}^{\mu}_a = - {\hat v}^{\nu}_a
{{\partial {\hat v}^{\mu}_a}\over {\partial x_0^{\nu}}} .}
Now, recalling again that ${\hat v}$ vanishes at the punctures, it is
easy to verify that:
\eqn\luno{\langle I_{\hat{v}_a}
\sigma_1^{(0)}\rangle_{{\hat \eta}_0}
= {\rm tr}\left(\epsilon \Ma d_p\Mai\right)\ . }
In the following it will be convenient to choose complex coordinates, thus
we introduce  the isothermal
coordinates ${\x0}^{\mu} = (Z , \bar Z)$ and the local ones
$x^{\mu}_a = (\za, \zabar)$. The relation between $(Z , \bar Z)$
and $(\za, \zabar)$
involves the Beltrami differentials $\bel{a}$ and
$\bbel{a}$:
\eqn\isothermal{ dZ \otimes d\bar Z = |\If{a}|^2 (d\za +
\bel{a}d\zabar)\otimes (d\zabar + \bbel{a} d\za).}
Now the matrix $\Ma$ is written
\eqn\transition{ (\Ma)^{\mu}_{\,\,\nu} =
\left( \matrix{\If{a} & \If{a} \bel{a} & \cr
 \bIf{a} \bbel{a} & \bIf{a} & \cr}\right).}
Inserting this expression  into \luno\
and taking into account that $\hat{\eta}_0$ in complex
coordinates corresponds to the matrix
$\left(\matrix{0 & 1 \cr 1 & 0 \cr}\right)$ we obtain:
\eqn\cofinal{{\rm tr}\left(\epsilon d_p\Ma\Mai\right)
 = - i\left(d_p\log {\If{a}\over \bIf{a} } +
{\bel{a} d_p \bbel{a} -\bbel{a} d_p\bel{a}
\over 1-|\bel{a}|^2}\right) . }

We are now in condition to analyse a particular example.
The simplest non-trivial one is the vacuum expectation value of
$\sigma^{(0)}(P)$ when the world-sheet is the sphere with four fixed
points.  We identify the sphere with the compactified complex plane and
the four fixed points with $ 0, 1, P,\infty$.

Our world-sheet is characterised by a single complex modulus $m$ that
identifies the point $P$ in the isothermal frame. Thus,
the moduli space coincides with the complex plane with the three
singular points $0,1$ and $\infty$ taken out.
It can be covered by three open disks centred around the three
singularities. It is therefore clear that each disk is not simply connected
since its centre is removed.
Let us consider for example the disk centred at the origin. When $m$
belongs to this disk the map $Z(z_0, \bar z_0, m )$ relating the isothermal
coordinates to the background ones can be chosen according to the equation:
\eqn\plufix{Z(z, \bar z, m )\equiv z \left({m\over z_{P}}
\right)^{\theta_R (1-
|z|^2)}.}
$\theta_R(x)$ is a regularized  $C^{\infty}$ step function,
interpolating smoothly between $0$ and $1$:
\eqn\stfun{\theta_R(x) =\cases{0, & if $x\le 0$;\cr 1, &if $x\ge 1-
C$,\cr}}
with $0<C<1$ and $z_{P}$  is the coordinate of $P$.

Without any loss of generality we can choose for example $z_P=1/2 +
i\sqrt{2}/2$ and $C=3/4$.   With this choice it is easy to see that:
\eqn\condi{Z(z_i, \bar z_i, m )=z_i\quad {\rm for}\quad z_i= 0,\, 1,\, \infty\
,}
and:
\eqn\condj{Z(z_P, \bar z_P, m )=m\ .}
It is also easy to verify that  for $0<|m|<1$, $Z(z, \bar z, m )$ defines a
quasi-conformal map of the sphere with three fixed points into itself.
Furthermore, comparing with \transition\ one has:
\eqn\vanbel{\bel{0}(z_P;m) \equiv 0}
in the whole disk and
\eqn\factp{\If{0}(z_P;m) = \log { m\over z_P}}
in the disk without the origin.
It is clear that a completely analogous construction can be repeated for the
disks centred at $1$ and $\infty$ obtaining in particular:
\eqn\vanbell{\mu_i(z_P;m) \equiv 0\ ,} for $i=1,\infty$
and
\eqn\factpp{\If{1}(z_P;m) = \log { m-1\over z_P-1} \quad
{\rm and}\quad\If{\infty}(z_P;m) = \log { m-\half\over z_P-\half}\ .}
Now we see that $\langle\sigma^{(0)}(P)\rangle$
vanishes as a function of $m$ in every disk. Indeed the terms
of $\sigma^{(0)}(P)$ depending on $c$ and
$\gamma$ do not contribute since these fields have no sources, while the
last term in \observables\ vanishes due to \vanbell.
Thus, apparently,  the presence of a singularity in the center of each disk
does not affect the functional integral.

However, computing the discontinuity of $\langle\sigma^{(0)}(P)\rangle$
between two and more neighbouring disks we have, from \piff, \dimone,
\boh, \luno\ and \cofinal,
$$\bigl(\langle\sigma^{(0)}(P)\rangle\bigr)_{a,b}=
{1\over 2\pi }d_p\,  {\rm Im}\, \log{\lambda_a(z_P,m)\over
\lambda_b(z_P,m)}\ ,$$
and
$$ \bigl(\langle\sigma^{(0)}(P)\rangle\bigr)_{a,b,c}=0\ ,$$
where $a,\, b,\, c\,$ correspond to the three disks around
$0, 1$ and $\infty$.
Then, using \integralchain\ we verify that the value of the
vacuum expectation value of $\sigma^{(0)}(P)$ is $1$.

\newsec{Conclusions}

It is rather clear from this analysis that the singular points of the
moduli space, while not affecting the contributions of their
neighbourhood to the physical expectation value, determine the
discontinuities from which the final, non-trivial result gets its
origin.

It is also remarkable that this result implies a breakdown of the
Slavnov-Taylor identity since, the $s$-cohomology being trivial, if
this identity where verified, all the physical correlators should
vanish. It results clearly from our study how the singularities of the
moduli space have determined this breakdown.
One might speculate about the correspondence between this mechanism and the
instabilities discussed by Fujikawa in \ref\fuji{\FUJI}.

The third important point that has to be noted is that our explicit
calculation has been possible since we have been able to include
the whole neibourhood of each singularity into  a finite number of cells of the
moduli space. In this way these singularities do not correspond to any
divergence but determine the lack of global definition of the local correlators
which therefore correspond to non-trivial elements
of \v{C}ech-De Rham cohomology.
In our opinion this situation is suggestive of an analogous possibility
in the case of gauge theories
\bigskip
{\bf Acknowledgements}

A great part of the new ideas contained in this report have originated
from many discussions, in particular with R. Stora, during the stay of
C. B. at the Research Institute for Mathematical Science of Kyoto
University.  This work is partially supported by the ECPR, contract
SC1-CT92-0789.

\footatend\vfill\supereject\immediate\closeout\rfile\writestoppt
\baselineskip=14pt\centerline{{\bf References}}\bigskip{\frenchspacing%
\parindent=20pt\escapechar=` \input refs.tmp\vfill\eject}\nonfrenchspacing

\bye